\documentclass[
  aip,
  apl,
  reprint,
  preprintnumbers,
  floatfix,
  amsmath,
  amssymb,
  groupedaddress,
]{revtex4-1}

\usepackage[dvips]{graphicx}
\usepackage{amsmath}
\usepackage{times}
\usepackage{graphicx}% Include figure files
\usepackage{dcolumn}% Align table columns on decimal point
\usepackage{bm}% bold math
\usepackage{color}
\usepackage{sidecap}
\usepackage{amssymb}
\usepackage{bbm}
\usepackage{mathrsfs}
\usepackage[sans]{dsfont}
\usepackage{epsfig,psfrag}
\usepackage{amsmath,amsfonts,amssymb}
\def\be{\begin{equation}}
\def\ee{\end{equation}}
\def\bea{\begin{eqnarray}}
\def\eea{\end{eqnarray}}

\begin{document}

\preprint{draft}

\title{Geometrical Phase Transition on WO$_3$ Surface}
\author{Abbas Ali Saberi}\email{a$_$saberi@ipm.ir \& ab.saberi@gmail.com }
\address {School of Physics, Institute for Research in Fundamental
Sciences (IPM), P.O.Box 19395-5531, Tehran, Iran}
\date{\today}

\begin{abstract}
A topographical study on an ensemble of height profiles obtained
from atomic force microscopy techniques on various independently
grown samples of tungsten oxide WO$_3$ is presented by using ideas
from percolation theory. We find that a continuous 'geometrical'
phase transition occurs at a certain critical level-height
$\delta_c$ below which an infinite island appears. By using the
finite-size scaling analysis of three independent percolation
observables i.e., percolation probability, percolation strength and
the mean island-size, we compute some critical exponents which
characterize the transition. Our results are compatible with those
of long-range correlated percolation. This method can be generalized
to a topographical classification of rough surface models.
\end{abstract}


\maketitle

The growth of rough surfaces and interfaces with many of their
scaling and universality properties has attracted the attention of
statistical physicists over the last three decades \cite{stanley}.
One of the less studied subjects is the one concerned with the
topographical properties of the self-affine surfaces. In this
letter, we present a systematic investigation of percolation
transition on an ensemble of experimentally grown WO$_3$ surfaces.
This gives some 'geometrical' characteristic exponents which well
describe a geometrical phase transition at a certain level-height
below which an infinite cluster-height (or island) appears. The
results of our work are closely related to the statistical
properties of the size distribution of the islands and their
coastlines together with their fractal properties.

The tungsten oxide WO$_3$ surface is one of the most interesting
metal oxides. It has been investigated extensively because for its
distinctive applications such as electrochromic
\cite{Granqvist,Bueno,Azimirad,Kuai,Takeda}, photochromic
\cite{Avellaneda}, gas sensor \cite{Kim,György,Kawasaki},
photo-catalyst \cite{Gondal}, and photoluminescence properties
\cite{Feng}. Many properties of the WO$_3$ thin films are related to
their surface structure as well as to their surface topography and
statistics such as grain size and height distribution. These
properties can also be affected during the growth process by either
deposition method or imposing external parameters on the grown
surfaces such as annealing temperature \cite{saberi}.

\begin{figure}[b]\begin{center}
\epsfxsize=5.8 truecm\center\epsfbox{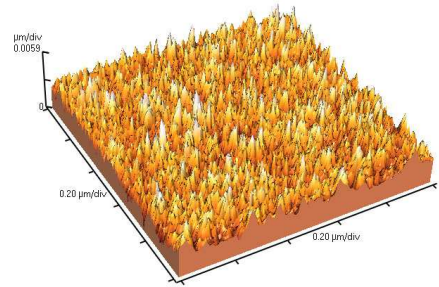}
 \narrowtext \caption{\label{Fig1}(Color
online) AFM image of WO$_3$
 thin film in scale 1 $\mu$m $\times$1 $\mu$m with resolution of 1/256 $\mu$m.}\end{center}
\end{figure}

In order to have an ensemble of WO$_3$ height profiles, 34 samples
were independently, and under the same conditions, deposited on
glass microscope slides with an area 2.5 cm$\times$2.5 cm using
thermal evaporation method. The deposition system was evacuated to a
base pressure of $\sim$ 4$\times$10$^{-3}$ Pa. The thickness of the
deposited films was chosen to be about 200 nm, and measured by
stylus and optical techniques. Using atomic force microscopy (AFM)
techniques, we have obtained 300 height profiles from the grown
rough surfaces with resolution of 1/256 $\mu$m in the scale of 1
$\mu$m$\times$1 $\mu$m (an example is shown in Fig. \ref{Fig1}). The
AFM scans were performed in various non-overlapping domains (10
images from each sample) from the centric region of the deposited
samples in order for the profiles to be statistically independent.

%%%%%%%%%%%%%%%%%%%%%%%%%%%%%%%%%%%%%%%%%%%%%%%%%%%%%%%%%%%%%%%%%%%%%%%%%%%%%%%%%%%%%%%%%%%%%%%%%%%%%%%%%%%%%%%%
%%%%%%%%%%%%%%%%%%%%%%%%%%%%%%%%%%%%%%%%%%%%%%%%%%%%%%%%%%%%%%%%%%%%%%%%%%%%%%%%%%%%%%%%%%%%%%%%%%%%%%%%%%%%%%%%%%%%

The percolation problem \cite{sahimi,SA} is an example of the
simplest pure geometrical phase transitions with nontrivial critical
behavior, and it is closely related to the surface topography. Let
us suppose a sample of height profile $\{h(\textbf{r})\}$ is a
topographical landscape. Now, let us imagine flooding this landscape
and coloring the parts above the water level white and the rest
black. If the water level is high, there will be small disconnected
islands, while if it is low, there will be disconnected lakes. There
is however a critical value of the sea level for which there is one
large supercontinent and one large ocean. As long as the original
height profile has a gaussian distribution with only short-range
correlations, it is believed that the large-scale properties of the
coastlines correspond to standard percolation cluster boundaries and
thus should be described by the theory of Schramm-Loewner evolution
(SLE) \cite{schramm} (see \cite{SLE} for a review of SLE). However,
the height profiles of WO$_3$ surface are quite different due to the
relevant contribution of long-range correlations \cite{Halperin}.
Hence, the corresponding coastlines of the height profiles belong
statistically to a different universality class \cite{Saberi}.

To define islands on a WO$_3$ surface, a cut is made at a certain
height $h_\delta=\langle h(\textbf{r})\rangle
+\delta\sqrt{\langle[h(\textbf{r})-\langle
h(\textbf{r})\rangle]^2\rangle}:=0$, where the symbol
$\langle\cdot\cdot\rangle$ denotes spatial averaging. Each island
(cluster-height) is then defined as a set of nearest-neighbor
connected sites of positive height. We show that there is a critical
level height denoted by the dimensionless parameter
$\delta=\delta_c$, at which a continuous percolation transition
occurs.

%%%%%%%%%%%%%%%%%%%%%%%%%%%%%%%%%%%%%%%%%%%%%%%%%%%%%%%%%%%%%%%%%%%%%%%%%%%%%%%%%%%%%%%%%%%%%%%%%%%%%%%%%%%%%%%%
\begin{figure}[t]\begin{center}
\includegraphics[scale=0.32]{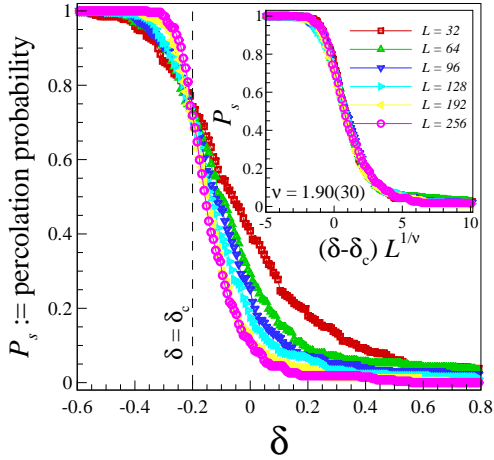}
\narrowtext\caption{\label{Fig2}(Color online) Main: probability
$P_s$ for the presence of a spanning island as a function of
$\delta$ measured for different lattice sizes $L$. The curves cross
at a critical level height $\delta_c=-0.20(1)$. Inset: data collapse
for the $P_s$ curves of different $L$ with $\nu=1.90$ and
$\delta_c=-0.20$.}\end{center}
\end{figure}
%%%%%%%%%%%%%%%%%%%%%%%%%%%%%%%%%%%%%%%%%%%%%%%%%%%%%%%%%%%%%%%%%%%%%%%%%%%%%%%%%%%%%%%%%%%%%%%%%%%%%%%%%%%%%%%%

To study the finite-size scaling (FSS) properties and measure
percolation observables which characterize the critical behavior, we
consider a box of dynamic size $L\times L$ from the centric region
of each original AFM sample. For each lattice size $16\leq L\leq
256$, each percolation observable was averaged over all number of
AFM samples.\\The first quantity we measure is the probability $P_s$
that at each level height $\delta$, an infinite island spans two
opposite boundaries of the box in just a specific direction, say
\emph{y}-direction. Ideally, the curves obtained for different
lattice sizes cross at a single point, marking the critical level
height $\delta_c$. As shown in Fig. \ref{Fig2}, the measured curves
cross at $\delta=\delta_c$, implying that the scaling dimension of
the percolation probability $P_s$ is zero.\\According to scaling
theory \cite{BH}, one expects that all the measured curves should
obey the scaling form \be\label{P_s}
P_s(\delta)=P_s[(\delta-\delta_c)L^{1/\nu}],\ee where the exponent
$\nu$ characterizes the divergence of the correlation length $\xi$
(proportional to the spatial extent of the islands) near the
percolation threshold, $\xi\sim |\delta-\delta_c|^{-\nu}$.\\We
measure the values of the exponent $\nu$ and the crossing point of
the curves $\delta_c$ by utilizing the data collapse. The quality of
the collapse of the curves is measured by defining a function
$S(\nu,\delta)$ of the chosen values $\nu$ and $\delta$ (the smaller
$S$ is indicative of a better quality of the collapse $-$ see
\cite{Kawashima, Bhattacharjee} and the appendix of \cite{HH}). We
find its minimum $S_{min}\sim1.78$ for $\nu=1.90(30)$ and
$\delta_c=-0.20(1)$. Inset of Fig. \ref{Fig2} illustrates the
collapse of all the $P_s$ curves, within the achieved accuracy, onto
a universal function by using the estimated values for $\nu$ and
$\delta_c$.

%%%%%%%%%%%%%%%%%%%%%%%%%%%%%%%%%%%%%%%%%%%%%%%%%%%%%%%%%%%%%%%%%%%%%%%%%%%%%%%%%%%%%%%%%%%%%%%%%%%%%%%%%%%%%%%%
\begin{figure}[b]\begin{center}
\includegraphics[scale=0.32]{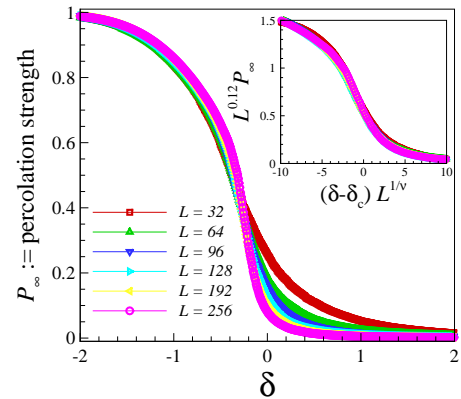}
\narrowtext\caption{\label{Fig3}(Color online) The percolation
strength $P_\infty$ of the islands as a function of $\delta$ for
different lattice sizes $L$. Inset: collapse of the data in the
suitably rescaled scales.}\end{center}
\end{figure}
%%%%%%%%%%%%%%%%%%%%%%%%%%%%%%%%%%%%%%%%%%%%%%%%%%%%%%%%%%%%%%%%%%%%%%%%%%%%%%%%%%%%%%%%%%%%%%%%%%%%%%%%%%%%%%%%

Percolation strength $P_\infty$, which measures the probability that
a point on a level height $\delta$ belongs to the largest island (or
equivalently, the fraction of sites in the largest cluster), is
another quantity defined as order parameter in the percolation. As
shown in Fig. \ref{Fig3}, we have computed $P_\infty$ as a function
of $\delta$. We find that the data follows the scaling form
\be\label{P_inf}P_\infty(\delta)=L^{-\tilde{\beta}}P_\infty[(\delta-\delta_c)L^{1/\nu}].
\ee Our best estimate for the exponent $\tilde{\beta}$
($\tilde{\beta}=\beta/\nu$ in percolation theory) is
$\tilde{\beta}=0.12(4)$ which was obtained by optimizing the quality
of the collapsed data. The data collapse is shown in the inset of
Fig. \ref{Fig3}.

%%%%%%%%%%%%%%%%%%%%%%%%%%%%%%%%%%%%%%%%%%%%%%%%%%%%%%%%%%%%%%%%%%%%%%%%%%%%%%%%%%%%%%%%%%%%%%%%%%%%%%%%%%%%%%%%
\begin{figure}[h]\begin{center}
\includegraphics[scale=0.4]{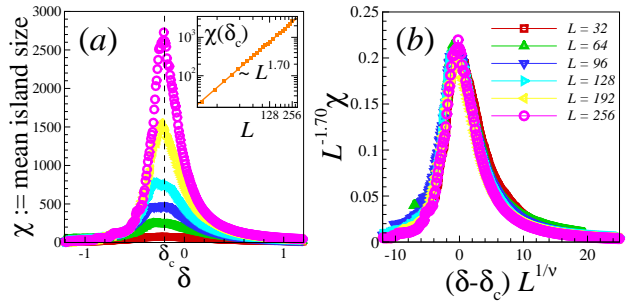}
\narrowtext\caption{\label{Fig4}(Color online) (a) mean island size
$\chi$ as a function of $\delta$ for different lattice sizes $L$.
Inset: $\chi$ at the critical level cut $\delta_c\simeq-0.20$ as a
function of $L$. (b) data collapse for $\chi$ for different lattice
sizes, according to Eq. (\ref{chi}).}\end{center}
\end{figure}
%%%%%%%%%%%%%%%%%%%%%%%%%%%%%%%%%%%%%%%%%%%%%%%%%%%%%%%%%%%%%%%%%%%%%%%%%%%%%%%%%%%%%%%%%%%%%%%%%%%%%%%%%%%%%%%%

Another independent observable is the mean island size
\be\label{MCZ}\chi=\frac{\sum'_s s^2 n_s}{\sum'_s s n_s},\ee where
$n_s$ denotes the average number density of islands of size $s$, and
the prime on the sums indicates the exclusion of the largest island
in each measurement.\\As presented in Fig. \ref{Fig4}, the obtained
curves $\chi(\delta)$ for different lattice sizes have their maximum
around the critical level $\delta_c$. We find that
$\chi(\delta_c)\sim L^{\tilde{\gamma}}$ with
$\tilde{\gamma}=1.70(3)$ ($\tilde{\gamma}=\gamma/\nu$ in percolation
theory)$-$ see inset of Fig. \ref{Fig4}(a). By using the exponents
$\tilde{\gamma}$ and $\nu$, it is possible to achieve a data
collapse according to the scaling form\be\label{chi}\chi(\delta)=
L^{\tilde{\gamma}}\chi[(\delta-\delta_c)L^{1/\nu}],\ee which is
shown in Fig. \ref{Fig4}(b).

A point which remains is determination of the universality class
which the observed percolation transition belongs to. The values we
find for the exponents are obviously not those of short-range
correlated percolation. In order to see whether they fit with
long-range correlated percolation, we calculate the two-point
correlation function $G(\textbf{x},\textbf{x}')$ which is defined as
the probability that two points of distance
$r=|\textbf{x}-\textbf{x}'|$, at the critical level height
$\delta_c$, belong to the same island. The best fit to our data
shows that $G$ is a decreasing function of the distance with a
power-law behavior $G(r)\sim r^{-\eta}$ and $\eta=0.34(2)$.

In the past, several papers have dealt with percolation with
long-range correlations \cite{Halperin,Coniglio,Isichenko,Prakash}
in which correlations \emph{decrease} with increasing $r$. A
completely different percolation with long-range correlations
\cite{Sahimi1} has also been proposed by Sahimi \cite{Sahimi2} in
which correlations \emph{increase} as $r$ does \cite{Sahimi3}. From
the above references, our results are in agreement with
\cite{Prakash} in which the value of $\nu$ is consistently lower
than the prediction $\nu=2/\eta$ given by \cite{Halperin}, for
$\eta\leq 1$ in two dimensions.\\In order to examine whether our
exponents satisfy well-known scaling and hyperscaling relations, we
have also computed the fractal dimension of the islands at the
critical level $\delta_c$, and found to be $d_f=1.83(3)$. It is thus
straightforward to see that our obtained exponents satisfy, within
the statistical errors, the following scaling relations for
$d=2$,\be
d_f=d-\tilde{\beta}=\frac{1}{2}(d+\tilde{\gamma})=\frac{1}{2}(d+2-\eta).\ee

In conclusion, analysis of independent percolation observables on an
ensemble of height profiles obtained by AFM images of WO$_3$
surfaces, revealed a continuous geometrical phase transition at a
certain critical level height $\delta_c$. We computed some critical
exponents which can be regarded as topographical characteristics of
the WO$_3$ surfaces. This method may lead to a topographical
classification of self-affine rough surfaces by computing the
exponents $\nu$, $\tilde{\beta}$, $\tilde{\gamma}$ and $\eta$ for
different growth models.

I would like to thank J. Cardy for his useful comments and M. Vincon
for critical reading of the manuscript. I also acknowledge financial
support from INSF grant.

\end{document}